\documentclass[aps,twocolumn,showpacs,preprintnumbers,superscriptaddress]{revtex4-2}
\usepackage{amsmath}
\usepackage{amssymb}
\usepackage{mathrsfs}
\usepackage{booktabs}
\usepackage{threeparttable} 
\usepackage{graphicx}
\usepackage{array}
\usepackage{float}
\usepackage{bm}
\usepackage[T1]{fontenc}
\usepackage{natbib}
\usepackage{braket}
\usepackage{epstopdf}
\usepackage{longtable}
\usepackage{multirow}
\usepackage{makecell}
\usepackage{siunitx}
\usepackage[colorlinks=true,citecolor=blue,linkcolor=blue,urlcolor=blue]{hyperref}
\DeclareUnicodeCharacter{2212}{-}

\begin{document}

\newcommand{\etal}{{\sl et al.}}
\newcommand{\ie}{{\sl i.e.}}
\newcommand{\mub}{$\mu_{\rm B}$}

\title{From trigonal to triclinic: Symmetry-tuned Rashba effects in buckled honeycomb SrHfO$_{3}$-based heterostructures}

\author{Okan K\"oksal}
\email{Corresponding author.\newline E-mail address: okan.koksal@Ugent.be}
\affiliation{Department of Electrical Energy, Metals, Mechanical Construction and Systems, Ghent University, Belgium} 
\affiliation{Center for Molecular Modeling, Ghent University, Belgium}

\date{\today}

\begin{abstract}
Harnessing the interplay of symmetry breaking and spin-orbit coupling, we investigate Rashba spin splitting in buckled honeycomb (SrHfO$_3$)$_2$/(LaAlO$_3$)$_4$(111) superlattices using density functional theory (DFT) calculations with a Hubbard $U$ term and a Wannier-based tight-binding (TB) model. In the non-centrosymmetric $P1$ phase, pronounced Rashba-type splitting emerges near the $M$ and $K$ points accompanied by a helical in-plane spin texture, while the centrosymmetric $P321$ phase remains spin-degenerate. A Wannier-based tight-binding Hamiltonian, extended analytically with on-site spin-orbit coupling, reproduces the DFT results. A Rashba coefficient of $\alpha_R = 0.34$\,eV$\,\cdot$\,\AA\ and energy $E_R = 29$\,meV are extracted directly from the DFT band structure placing the system among moderately strong oxide Rashba materials. $\Gamma$-phonon calculation confirms the dynamical stability of the $P1$ structure and the results reveal the critical role of symmetry breaking and inter-orbital hybridization in enabling Rashba effects, supported by enhanced imaginary second-nearest-neighbor hoppings and Berry curvature. These findings establish SrHfO$_3$-based buckled heterostructures as a promising platform for engineering Rashba effects in oxide-based spintronic devices.
\end{abstract}

\maketitle
\section{Introduction}
Complex oxide heterostructures have emerged as a versatile platform to explore novel states of matter arising from the interplay of lattice symmetry, strong electronic correlations and relativistic spin-orbit coupling (SOC)~\cite{Zutic2004, Bihlmayer2017}. A prominent example is the interface between the band insulators LaAlO$_3$ (LAO) and SrTiO$_3$ (STO), which exhibits a rich variety of phenomena including two-dimensional electron gases (2DEGs), superconductivity, magnetism and metal-insulator transitions~\cite{Ohtomo2004, Thiel2006, Bert2011, Sulpizio2014, Vonk2012, Cancellieri2011, Huijben2009, Mannhart2010, Pentcheva2010, Zubko2011, Pentcheva2012, Hwang2012, Reyren2007, Pentcheva2009, Doennig2013}. While most studies have focused on the (001) orientation, the (111) stacking sequence offers an alternative geometry where triangular and honeycomb lattice motifs naturally arise, hosting topologically nontrivial band structures and orbital-selective physics~\cite{Doennig2013, Xiao2011, Ruegg2011}. Recent theoretical and experimental works have highlighted the (111) interface as a platform for Dirac semimetals, multiferroics as well as charge-ordered insulators, depending on symmetry conditions and strain~\cite{Doennig2013, Herranz2012}. In the bilayer limit, the BO$_6$ octahedra form a buckled honeycomb lattice, analogous to graphene, with the potential for hosting massless Dirac fermions protected by inversion and rotation symmetry ($P321$). Once inversion symmetry is lifted, the system may undergo a transition to a phase with charge ordering and orbital polarization~\cite{Doennig2013}.

In parallel, the emergence of a switchable Rashba effect in oxide heterostructures has attracted significant interest for spintronic applications. The Rashba effect, which has been originally proposed by Bychkov and Rashba~\cite{Bychkov1984, Winkler2003, LaShell1996} results from broken inversion symmetry and enables electric control over spin precession, with far-reaching implications for Majorana physics and spin-orbitronics~\cite{Zutic2004}. One of the earliest proposals to exploit this effect was the spin field-effect transistor introduced by Datta and Das~\cite{Datta1990}, where spin precession is modulated via gate-controlled Rashba coupling. In oxide systems such as LAO/STO(001), the Rashba interaction plays a central role in stabilizing superconductivity~\cite{Caviglia2010, BenShalom2010} and may even reconcile coexisting superconducting and magnetic phases~\cite{Dikin2011, Michaeli2012}. At the same time, oxygen-deficient SrTiO$_3$ surfaces have been shown to exhibit coexisting magnetism, spin texture and in-gap states. These findings highlight how site-dependent contributions at the surface of oxygen-deficient SrTiO$_3$ contribute differently to the observed physical properties, which turns out to be a key factor for tuning surface spin-orbit phenomena~\cite{Altmeyer2016}. Heavy-element substrates such as KTaO$_3$ (KTO) have been shown to dramatically enhance SOC effects due to their large atomic number and intrinsic polarity. Notably, the study~\cite{Geisler2023} demonstrates that the Rashba splitting in NdNiO$_2$/KTO(001) can exceed 200\,meV, a magnitude that is comparable to Bi(111) and is attributed to interfacial polarity and orbital-selective effects. In contrast to more conventional perovskite interfaces, the use of KTO enables a cubic-to-linear transition in spin splitting and anisotropic Fermi surface reshaping. The Rashba constants in these systems are one to two orders of magnitude larger than those in LAO/STO(001)~\cite{Caviglia2010, Geisler2023, King2012}. These findings underscore the importance of interfacial polarity, strain, orbital character and substrate choice in engineering tunable Rashba systems. By combining heavy-element layers with ferroelectric distortions, it is possible to induce and control large spin splittings using electric fields~\cite{Zhong2015, King2012, Liu2021}. The Rashba energy and spin texture can thus be modulated through heterostructure design, symmetry breaking, strain and external gating.

In this work, the (SrHfO$_3$)$_2$/(LaAlO$_3$)$_4$(111) heterostructure under tensile strain imposed by the STO substrate ($a_{\mathrm{STO}}^{\mathrm{GGA}}\,=\,3.92$\,\AA)~\cite{Doennig2013} is investigated. Using density functional theory (DFT) calculations including GGA$+U$(+SOC), the electronic band structure and spin texture in both $P1$ (non-centrosymmetric) and $P321$ (centrosymmetric) symmetries are examined. A dispersive metallic band crossing the Fermi level, accompanied by a pronounced Rashba-type spin winding at the $M$ and $K$ points, is observed in the $P1$ phase. Conversely, the $P321$ symmetry suppresses inversion-breaking effects and the associated spin splitting. A tight-binding (TB) model based on maximally localized Wannier functions is constructed, which includes both inter-site hopping and intra-atomic SOC. This enables an analytical extraction of Rashba parameters such as the Rashba coefficient $\alpha_R$, Rashba energy $E_R$ as well as an investigation of their dependence on SOC strength. Moreover, the role of symmetry and orbital character is dissected by analyzing the effect of hopping matrices $t_1$, $t_2$ and evaluating the dynamical stability of the structure including SOC contributions. The study establishes the (SrHfO$_3$)$_2$/(LaAlO$_3$)$_4$(111) system as a viable oxide-based platform for realizing and tuning Rashba spin-split metallic states with implications for low-dimensional spintronics and oxide electronics.

\begin{figure}[htbp!]
\centering
\includegraphics[width=0.5\textwidth]{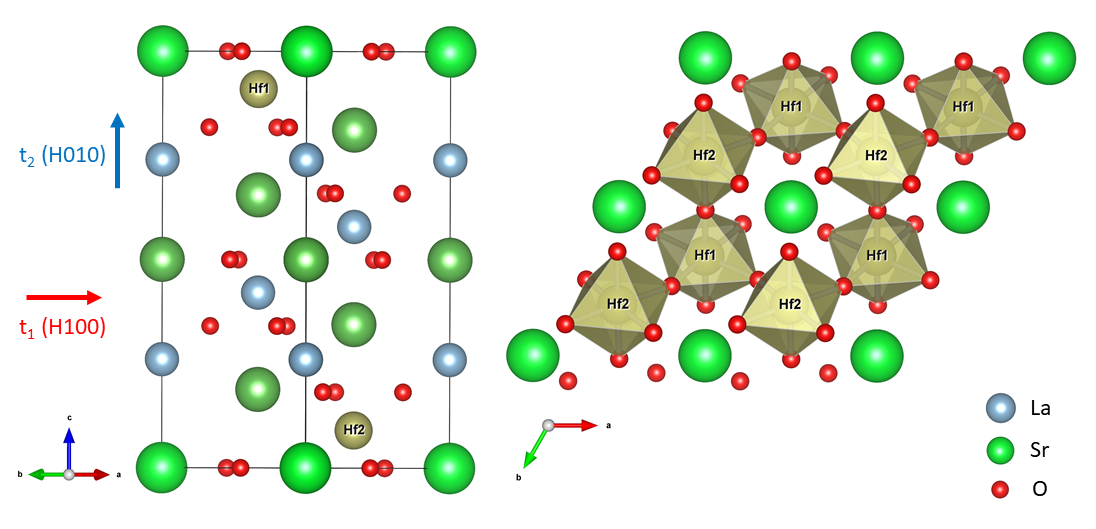}
\caption{Crystal structure of the (SrHfO$_3$)$_2$/(LaAlO$_3$)$_4$(111) buckled honeycomb superlattice. The left panel a) shows a side view along the $a$-axis, highlighting the layered stacking of La (blue), Sr (green), O (red) and Hf (beige) atoms. The $t_1$ and $t_2$ labels denote nearest-neighbor (red arrow) and second-nearest-neighbor (blue arrow) in-plane hopping paths within the buckled honeycomb layer. The right panel b) presents a top view along the $c$-axis emphasizing the two inequivalent Hf sites (Hf1 and Hf2) forming the distorted HfO$_6$ octahedra.}
\label{figure1}
\end{figure}

\section{Structure and computational methods}

The DFT+$U$ calculations for (SrHfO$_3$)$_2$/(LaAlO$_3$)$_4$(111) were carried out using the VASP code~\cite{VASP} with the projector augmented wave (PAW) method~\cite{Kresse1999}. The exchange-correlation functional was treated within the generalized gradient approximation (GGA) of Perdew, Burke and Ernzerhof (PBE). To account for static local electronic correlations, the GGA+$U$ approach was employed in the Dudarev formalism~\cite{Dudarev1998}, which uses an effective interaction $U_{\mathrm{eff}} = U - J$. For Hf $5d$ states, $U = 2$~eV and $J = 0.5$~eV were applied, while for La $4f$ states, a Coulomb repulsion term of $U = 8$~eV was used. The in-plane lattice constant was fixed to $a_{\mathrm{STO}} = 3.92$~\AA, corresponding to epitaxial strain imposed by an STO(111) substrate. The simulation cell of the perovskite-derived superlattice (see Fig.~\ref{figure1}) contains 30 atoms: 1 Sr, 5 La, 18 O, 4 Al, and 2 Hf cations. The simulations were performed using a $\Gamma$-centered $k$-point mesh of $12 \times 12 \times 2$ and a plane-wave energy cutoff of 600~eV. For hybrid functional calculations with HSE06~\cite{Heyd2003}, a reduced $k$-point mesh of $8 \times 8 \times 2$ was used. The lattice parameter $c$ and all atomic positions were relaxed until the residual Hellmann--Feynman forces on each atom were below 1~meV/\AA.

\section{Results and Discussion}

\subsection{Tight-Binding Hamiltonian}

Although Hf$^{4+}$ in SrHfO$_3$ is formally in a 5$d^0$ configuration, the conduction bands exhibit predominantly Hf 5$d_{xy}$ and 5$d_{xz}$ orbital character belonging to the $t_{2g}$ manifold. To analyze the Rashba spin splitting and electronic structure of the (SrHfO$_3$)$_2$/(LaAlO$_3$)$_4$(111) heterostructure, a tight-binding (TB) Hamiltonian based on maximally localized Wannier functions (MLWFs) projected onto the Hf $d_{xy}$ and $d_{xz}$ orbitals was constructed. The unit cell contains two inequivalent Hf atoms (Hf1 and Hf2), each contributing two orbitals, resulting in a four-band spinless Hamiltonian. While the well-known minimal lattice model such as the Kane-Mele model for graphene incorporate spin-orbit coupling via next-nearest-neighbor (NNN) terms to open a topological gap at the Dirac points \cite{KaneMele2005}, the present system involves a heavy atom (Hf) and broken inversion symmetry, where Rashba-type SOC arises more naturally. This Rashba-type mechanism complements prior analytical treatments of Rashba-type SOC in $d$-orbital systems under inversion-breaking fields~\cite{Shanavas2014}. Spin textures have previously been employed to characterize Chern insulating phases in oxide-based systems with strong spin-orbit coupling and broken inversion symmetry, such as in (La$X$O$_3$)$_2$/(LaAlO$_3$)$_4$(111) heterostructures and EuO/MgO(001) superlattices~\cite{Koksal2019SciRep,Koksal2021PRB, Guo2017NPJ}, emphasizing the critical role of spin-orbit coupling and band inversion mechanisms in driving topologically nontrivial electronic states. In contrast to the Kane-Mele model, which preserves spin degeneracy under time-reversal and inversion symmetry, our approach uses a material-specific tight-binding Hamiltonian derived from \textit{ab initio} Wannier functions and explicitly incorporates atomic SOC on-site, allowing us to capture realistic Rashba splitting effects. As the DFT calculation is nonmagnetic, the wannier90~\cite{Mostofi2008} result yields a spin-degenerate Hamiltonian $H_{\mathrm{W}}$ constructed over the energy window $[-0.8, 0.1]$ eV. 

To capture spin-orbit coupling (SOC) effects, the orbital basis was doubled and analytically SOC was introduced as an on-site atomic interaction of the form $\lambda \vec{L} \cdot \vec{S}$, coupling $d_{xy}^{\uparrow}$, $d_{xz}^{\downarrow}$ and $d_{xz}^{\uparrow}$, $d_{xy}^{\downarrow}$ channels. This yields an $8 \times 8$ spinor Hamiltonian of the form:

\begin{align}
H_0 &= H_W + H_{\mathrm{SOC}} \notag \\
    &= \sum_{i,j,s} t_{ij} \, c_{j,s}^\dagger c_{i,s} 
    + \sum_{i,j,s,s'} \lambda_{\mathrm{SOC}} \vec{L}_{ij} \cdot \vec{S}_{s s'} \, c_{j,s'}^\dagger c_{i,s}
\end{align}

where $t_{ij}$ are spinless hopping matrix elements obtained from Wannierization, $\vec{L}_{ij}$ encodes the orbital angular momentum structure and $\vec{S}$ are the spin Pauli matrices. This formalism allows for interpolation of SOC strength $\lambda$ without requiring further \textit{ab initio} calculations. The explicit SOC matrix in the spinor basis is given by:

\begin{equation}
H_{\mathrm{SOC}} = \lambda
\begin{pmatrix}
0 & 0 & 0 & 0 & 0 & -i & 0 & 0 \\
0 & 0 & 0 & 0 & i & 0 & 0 & 0 \\
0 & 0 & 0 & 0 & 0 & 0 & 0 & -i \\
0 & 0 & 0 & 0 & 0 & 0 & i & 0 \\
0 & i & 0 & 0 & 0 & 0 & 0 & 0 \\
-i & 0 & 0 & 0 & 0 & 0 & 0 & 0 \\
0 & 0 & 0 & -i & 0 & 0 & 0 & 0 \\
0 & 0 & i & 0 & 0 & 0 & 0 & 0
\end{pmatrix}
\end{equation}

with $\lambda = 0.1-0.15$ eV chosen in accordance with \textit{ab initio} estimates for the Hf $5d$ orbitals. This model successfully reproduces the band dispersion and Rashba spin splitting observed in the first-principles calculation, particularly around the $M$ and $K$ points. In the non-centrosymmetric $P1$ phase, the SOC-induced spin texture exhibits pronounced helical winding, characteristic of Rashba-type effects. Owing to its higher symmetry, the centrosymmetric $P321$ phase preserves inversion and threefold rotational symmetry, thereby suppressing spin splitting and leading to spin-degenerate bands.

Fig.~\ref{figure1} illustrates the atomic structure of the SrHfO$_3$/LaHfO$_3$(111) SL in a buckled honeycomb arrangement. The left view shows the stacking of alternating SrO and LaO layers along the out-of-plane direction with two inequivalent Hf sites labeled Hf1 and Hf2 embedded in octahedral oxygen cages. The $t_1$ and $t_2$ arrows represent in-plane hopping between Hf sites with $t_1$ connecting nearest neighbors and $t_2$ connecting next-nearest neighbors along the H(100) and H(010) directions. The alternating Hf1 and Hf2 polyhedra in the right view form the buckled honeycomb lattice, which hosts the relevant $d$-orbital manifold responsible for Rashba-type spin splitting in this system. This topology, combined with the local octahedral distortions and inequivalent chemical environments, breaks inversion symmetry and facilitates Rashba-type spin-orbit coupling effects. The identification of the dominant hopping pathways is essential for constructing minimal tight-binding models and understanding the origin of spin splitting and Berry curvature in this system.

\subsection{HSE06 and GGA + $U$ (+SOC) results}

\begin{figure*}[htbp!]
\centering
\includegraphics[width=1.0\textwidth]{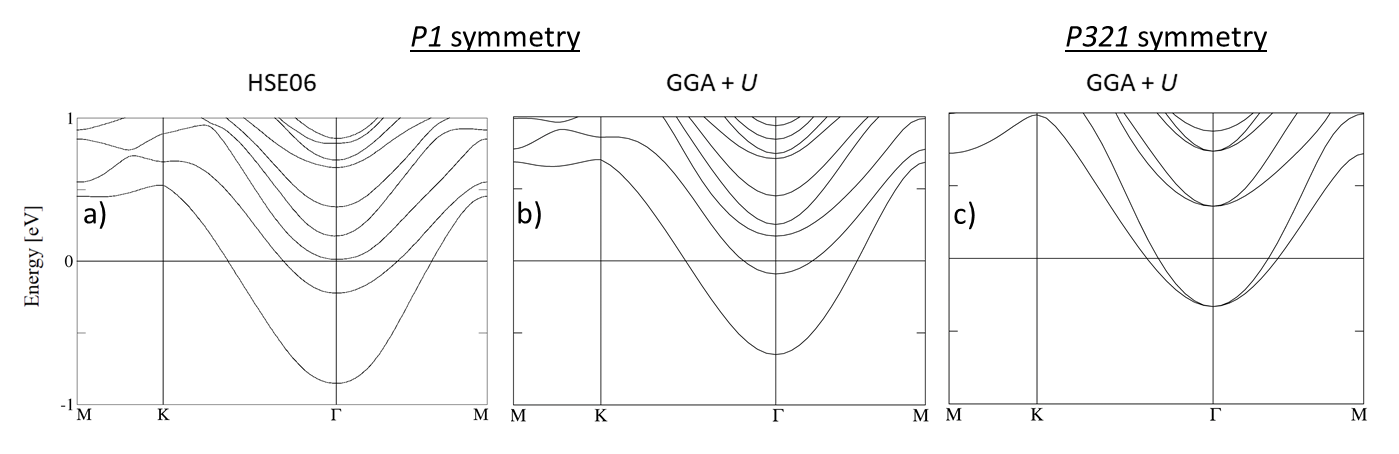}
\caption{Comparison of spin-degenerate band structures computed at different levels of theory and symmetry: a) HSE06, b) GGA + $U$ for \textit{P1} symmetry and c) GGA + $U$ for \textit{P321} symmetry.}
\label{figure2}
\end{figure*}

\begin{figure}[htbp!]
\centering
\includegraphics[width=0.5\textwidth]{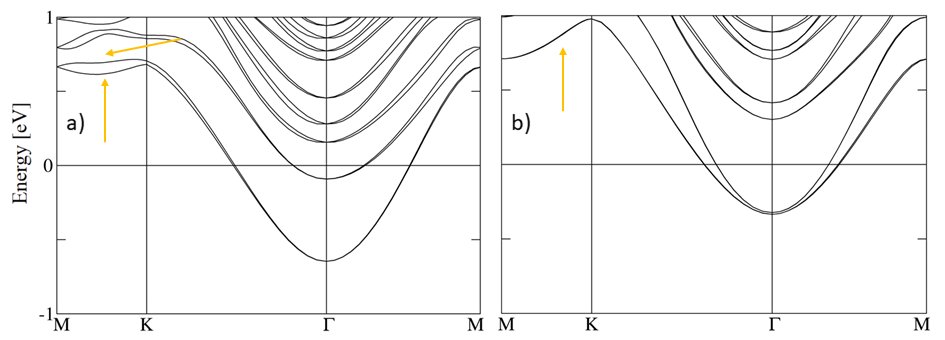}
\caption{Comparison of GGA+$U$+SOC band structures for the distorted a) $P1$ and symmetric b) $P321$ phases of SrHfO$_3$-based buckled honeycomb SL. The highlighted arrows indicate Rashba-type spin splittings along the $M$--$K$ direction. In the $P1$ phase, clear band splitting emerges due to strong spin-orbit coupling combined with inversion symmetry breaking, whereas in the $P321$ phase, symmetry constraints suppress inter-band hybridization and lead to weaker spin splitting.}
\label{figure3}
\end{figure}

\begin{figure*}[htbp!]
\centering
\includegraphics[width=0.9\textwidth]{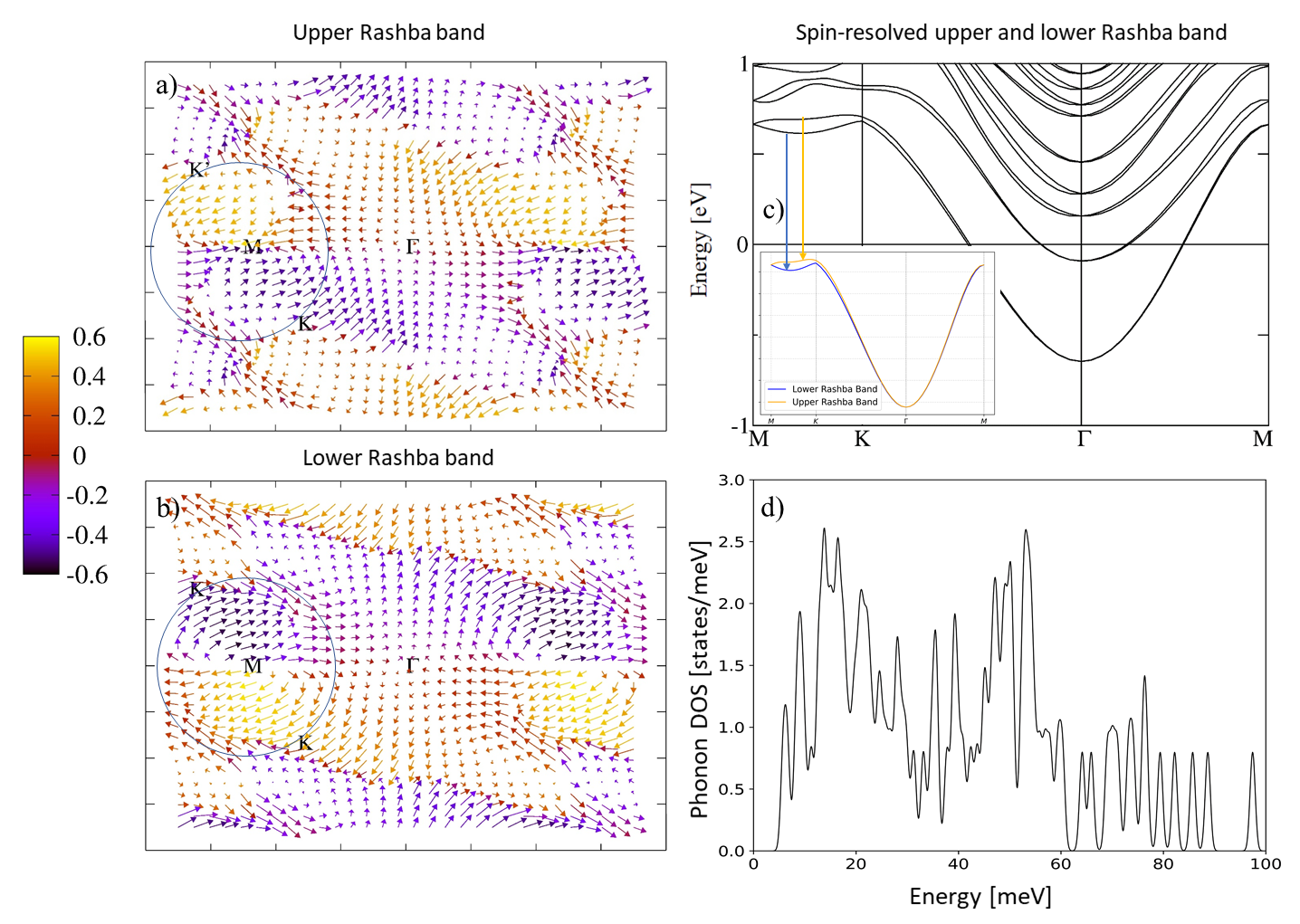}
\caption{Spin texture and band structure analysis of the Rashba-split conduction states in the $P1$ phase. a) Top-left and b) bottom-left panels show the spin-resolved textures of the upper and lower Rashba bands, respectively, in the $k_x$--$k_y$ plane, with arrow direction representing in-plane spin components and the color bar denoting the out-of-plane spin polarization $S_z$. The Rashba winding is prominent near the M point indicating strong spin-momentum locking and broken inversion symmetry in the low-symmetry configuration $P1$. c) The upper-right panel shows the DFT-derived band structure with Rashba splitting and its inset highlights the spin-resolved branches for the Rashba pair. d) Bottom-right panel displays the corresponding phonon density of states (DOS) confirming dynamical stability.}
\label{figure4}
\end{figure*}

The band structures shown for $P1$ symmetry using both HSE06 and GGA + $U$ methods demonstrate consistent features in the conduction bands near the Fermi level so that the GGA + $U$ result qualitatively reproduces the dispersion and curvature observed in the HSE06 calculation for the SrHfO$_3$-based buckled honeycomb lattice. With preserved $P321$ symmetry, the GGA + $U$ band structure retains the overall shape and conduction band minimum (CBM) position but exhibits enhanced degeneracy and reduced band warping due to the higher crystallographic symmetry. These differences are crucial for understanding spin-orbit coupling (SOC) effects, as the availability of orbital mixing channels and band curvature at high-symmetry points determine the strength and distribution of Berry curvature, as observed in subsequent Berry curvature analysis (see Fig.~\ref{figure7}b).

The band structure comparison emphasizes the role of symmetry in modulating Rashba spin splitting in SrHfO$_3$-derived SLs. The comparison between GGA\,+\,$U$ and GGA\,+\,$U$\,+\,SOC (cf. Fig.~\ref{figure3}a,b) reveals that spin-orbit coupling (SOC) lifts degeneracies near the $\Gamma$-point by introducing avoided crossings in otherwise spin-degenerate bands. In the $P1$ phase (cf. Fig.~\ref{figure3}a) which lacks inversion symmetry, the bands exhibit pronounced spin splitting along the $M$--$K$ direction, a hallmark of the Rashba effect. This is consistent with the presence of allowed inter-orbital SOC matrix elements and reduced symmetry constraints. In contrast, the $P321$ phase retains threefold rotational symmetry and inversion symmetry, which restrict the available SOC pathways. As a result, the spin-split features are significantly diminished, particularly near $K$ and where spin textures are symmetry-protected. This contrast corroborates that symmetry lowering from $P321$ to $P1$ enables stronger Rashba-type effects by enhancing spin-orbit-induced inter-band mixing.

\subsection{Spin texture and dynamical stability}

Fig.~\ref{figure4} presents a comprehensive analysis of the Rashba effect in the $P1$ symmetry phase. The spin texture plots (see Fig.~\ref{figure4}a,b) in the Brillouin zone reveal clear helicity for both the upper and lower Rashba bands, particularly centered around the $M$ point. The in-plane spin orientation exhibits the typical chiral structure expected from Rashba-type spin splitting, while the out-of-plane component $S_z$ shows modulations that underpin non-trivial spin-orbit coupling contributions. The band structure panel confirms the presence of Rashba-split bands near the conduction band minimum, whereas the inset verifies that these are spin-resolved branches of the same band pair (see Fig.~\ref{figure4}c). Finally, the phonon density of states (DOS) as displayed in Fig.~\ref{figure4}d verifies the vibrational stability of the structure, ensuring that the observed spin texture and band features correspond to a dynamically stable electronic configuration. Importantly, the absence of imaginary phonon modes in the vibrational spectrum provides clear evidence for the dynamic stability of the $P1$ phase as a local minimum. The results emphasize that spin-orbit coupling in this distorted phase produces both in-plane spin winding and out-of-plane polarization as a consequence of strong inversion symmetry breaking in $P1$. The phonon spectrum was obtained from a $\Gamma$-point calculation including spin-orbit coupling (DFT\,+\,$U$\,+SOC) confirming that the $P1$ phase remains dynamically stable even when SOC is explicitly taken into account. Not only is the $P1$ phase dynamically stable, but it is also energetically preferred over the $P321$ configuration by 0.12~eV based on GGA+$U$(+SOC) total energies. This energetic stabilization further supports the spontaneous symmetry breaking toward the non-centrosymmetric phase under tensile strain. Furthermore, attempts to stabilize alternative spin arrangements, i.e., ferromagnetic and antiferromagnetic orderings, consistently relaxed back to the nonmagnetic solution indicating that the Rashba-active $P1$ phase is the ground state in terms of magnetic configuration.

\begin{figure*}[htbp!]
\centering
\includegraphics[width=1.0\textwidth]{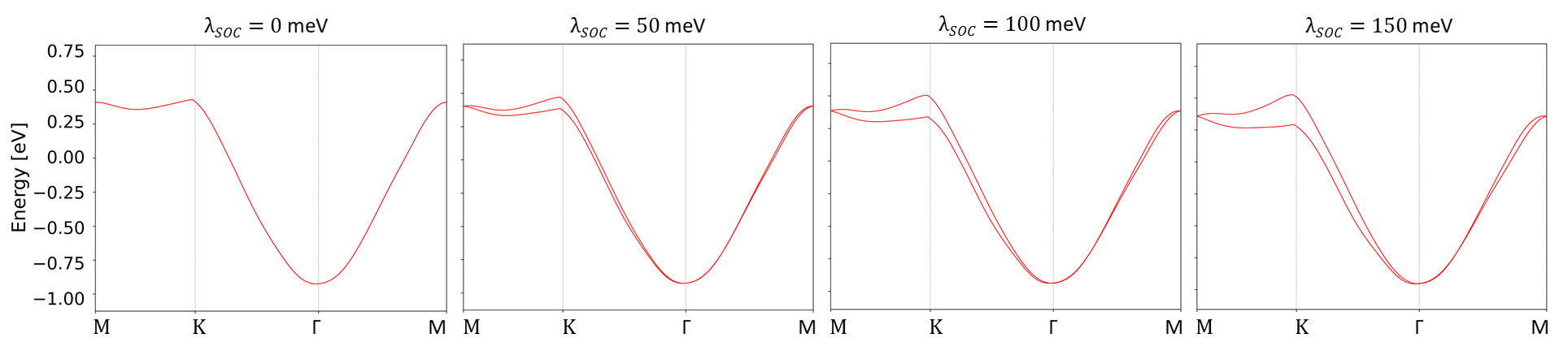}
\caption{Evolution of the two-band tight-binding model as a function of spin-orbit coupling strength $\lambda_{\mathrm{SOC}}$. At $\lambda_{\mathrm{SOC}} = 0$ meV, the two bands are spin-degenerate and display no visible splitting. With increasing $\lambda_{\mathrm{SOC}}$, Rashba-type band splitting emerges and intensifies near the $K$ point and along the $M \rightarrow K$ direction. This progressive band deformation highlights the role of SOC in lifting band degeneracies and inducing spin-momentum locking, reflecting symmetry-allowed Rashba effects in the $P1$ phase.}
\label{figure5}
\end{figure*}

Fig.~\ref{figure5} illustrates the progressive development of Rashba-type spin splitting in a two-band tight-binding model as a function of increasing spin-orbit coupling (SOC) strength. For $\lambda_{\mathrm{SOC}} = 0$~meV, the bands are spin-degenerate and exhibit no SOC-related features. Upon introducing SOC, starting at $\lambda_{\mathrm{SOC}} = 50$~meV, a subtle but distinct lifting of degeneracies becomes apparent near non-time-reversal-symmetric $k$-points such as $K$. This behavior signals the onset of Rashba-like spin splitting, where bands begin to separate due to spin-momentum locking. With further increase to $\lambda_{\mathrm{SOC}} = 100$~meV and $150$~meV, the spin splitting intensifies and extends across a wider region of the Brillouin zone. The gap between the Rashba-split branches enlarges and the band curvature reflects stronger SOC-induced mixing between spin channels. These features originate from the interplay of broken inversion symmetry and finite SOC, which enable off-diagonal spin-mixing terms in the Hamiltonian. Overall, this evolution highlights the critical role of SOC in shaping the band structure and spin texture. The model captures essential mechanisms driving Rashba physics and demonstrates how tuning $\lambda_{\mathrm{SOC}}$ controls the emergence of spin-split states, relevant for spintronics as well as topological phenomena.

\subsection{Rashba parameter and analysis of hopping amplitudes}

\begin{figure}[htbp!]
\centering
\includegraphics[width=0.48\textwidth]{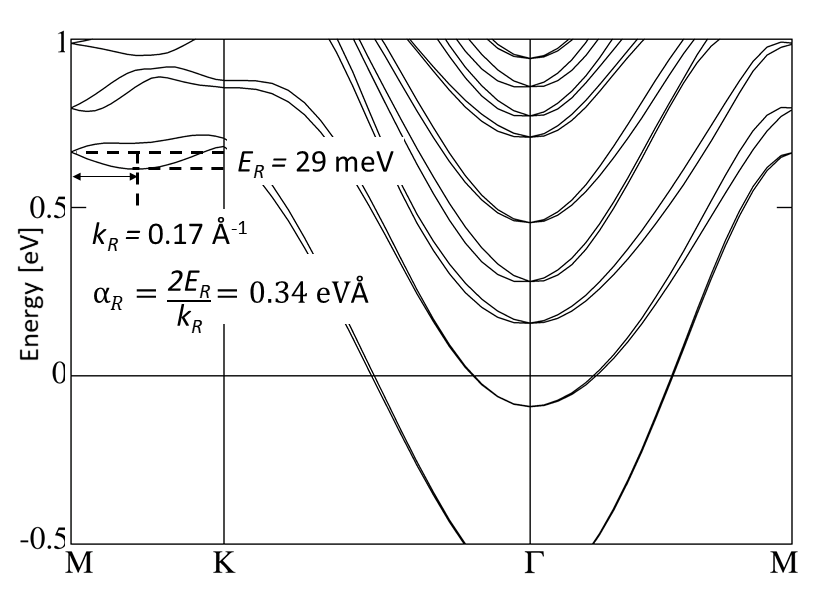}
\caption{Direct extraction of the Rashba parameters from the DFT+$U$+SOC band structure in the $P1$ phase along the $M$--$K$ path. The Rashba energy $E_R = 29$\,meV and momentum offset $k_R = 0.17$\,\AA$^{-1}$ yield a Rashba coefficient of $\alpha_R = 2E_R / k_R = 0.34$\,eV$\cdot$\AA. The inset highlights the spin-split conduction band pair with the extracted quantities annotated.}
\label{figure5_1}
\end{figure}

To quantify the spin-orbit-induced band splitting in the $P1$ phase, we extract the Rashba energy $E_R$ and momentum offset $k_R$ directly from the DFT+$U$+SOC band structure along the $M$--$K$ direction, as shown in Fig.~\ref{figure5_1}. The Rashba energy $E_R$ is defined as the energy difference between the band degeneracy point (where the spin-split branches would cross in the absence of SOC) and the extremum (typically a minimum) of the spin-split bands. In this system, $E_R$ corresponds to the vertical energy offset between the band crossing point and the bottom of the lower spin-split branch. The momentum offset $k_R$ denotes the horizontal distance in $k$-space between the high-symmetry $M$ point and the minimum of the lower spin-split conduction band. From the annotated band structure, we obtain $E_R = 29$\,meV and $k_R = 0.17$\,\text{\AA}$^{-1}$ (cf. Fig.~\ref{figure5_1}), resulting in a Rashba coefficient of $\alpha_R = 2E_R / k_R = 0.34$ eV $\cdot$\AA. These values position the system in the moderate Rashba regime. While $\alpha_R$ is sizable, the relatively small $E_R$ and $k_R$ reflect a moderate rather than strong Rashba effect, in line with the classification of Zunger \textit{et al.}~\cite{mera2020rashba}. This approach avoids assumptions about band curvature and instead extracts the Rashba parameters directly from the spin-split DFT bands, providing an intuitive and model-independent characterization of the Rashba effect in the $P1$ phase.

To further validate this result, we employed a tight-binding (TB) Hamiltonian derived from maximally localized Wannier functions and evaluated the Rashba coefficient using a $k \cdot p$ theory-based fit to the two lowest conduction bands along the $M$--$K$ path. At a fixed spin--orbit coupling strength of $\lambda_{\rm SOC} = 0.15$\,eV, this model yields a Rashba coefficient of $\alpha_R = 0.33$\,eV$\cdot$\text{\AA}, which is in excellent agreement with the direct DFT-based extraction. The two-band dispersion used for the fit follows from the diagonalization of a minimal $2 \times 2$ Rashba Hamiltonian of the form:

\begin{equation}
H(\mathbf{k}) = -\sigma_0 \frac{\hbar^2 k^2}{2m^*} + \alpha_R (k_y \sigma_x - k_x \sigma_y),
\end{equation}

where $m^*$ is the effective mass, $\alpha_R$ the Rashba coefficient, $\sigma_{x,y}$ the Pauli matrices and $\mathbf{k} = (k_x, k_y)$ the in-plane momentum.

Here, $\sigma_0$ denotes the $2 \times 2$ identity matrix and its presence with a negative prefactor ensures that the band minimum lies below the reference energy (e.g., the Fermi level), which is in line with the numerical TB-derived conduction band energies. The corresponding eigenvalues are:

\begin{equation}
E_\pm(k) = -\frac{\hbar^2 k^2}{2m^*} \pm \alpha_R k,
\end{equation}

describing a pair of spin-split parabolic bands with momentum-offset extrema. This model reproduces the Rashba-like dispersion along the $M$--$K$ path and confirms that the low-energy spin splitting can be accurately captured by an effective two-band theory. The good agreement between the direct DFT-based and model-based $\alpha_R$ values supports the use of minimal $k \cdot p$ Hamiltonians to describe and tune spin-orbit coupling effects in SrHfO$_3$-based heterostructures.

\begin{figure*}[htbp!]
\centering
\includegraphics[width=0.9\textwidth]{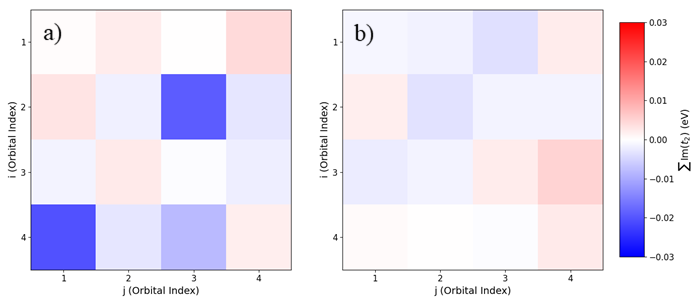}
\caption{Heatmaps showing the summed imaginary part of second-nearest-neighbor $t_2$ hopping amplitudes for a) $P1$ (left) and b) $P321$ (right) symmetries, resolved by orbital pair index $(i,j)$. The color scale indicates the magnitude and sign of $\mathrm{Im}(t_2)$ in units of eV. The $P1$ phase displays stronger and more asymmetric imaginary components, indicative of spin-orbit-driven inter-orbital mixing due to broken inversion symmetry. By comparison, the $P321$ structure exhibits more symmetric and suppressed values corresponding to preserved inversion and reduced Rashba activity.}
\label{figure6}
\end{figure*}

\begin{table}[htbp!]
\centering
\caption{Comparison of representative tight-binding parameters ($t_1$, $t_2$, $t'$, etc.) and on-site energies in units of eV between the $P1$ and $P321$ phases of the (SrHfO$_3$)$_2$/(LaAlO$_3$)$_4$(111) heterostructure. Shown are selected real-valued intra- and inter-orbital hoppings as well as diagonal on-site energies along symmetry-distinct $R$-vectors.}
\label{tab:P1_P321_TB_Comparison}
\begin{threeparttable}
\begin{tabular}{llccl}
\toprule
Parameter & Orbital pairing & $R$-vector & $P1$ [eV] & $P321$ [eV] \\
\midrule
$t_1$               & $\langle d_{xy} | \widehat{H} | d_{xy} \rangle$   & $\pm(1,0,0)$      & $-0.14$ & $-0.16$ \\
$t_2$               & $\langle d_{xz} | \widehat{H} | d_{xz} \rangle$   & $\pm(1,0,0)$      & $-0.14$ & $-0.16$ \\
$t_2^\text{diag}$   & $\langle d_{xy} | \widehat{H} | d_{xy} \rangle$   & $\pm(1,1,0)$      & $-0.11$ & $-0.17$ \\
$t'$                & $\langle d_{xz} | \widehat{H} | d_{xy} \rangle$   & $(-1,-1,0)$       & $ 0.03$ & $ 0.01$ \\
$t''$               & $\langle d_{xz} | \widehat{H} | d_{xy} \rangle$   & $(1,0,0)$         & $-0.01$ & $-0.01$ \\
\midrule
$\varepsilon_{d_{xz}}^{(1)}$ & $\langle d_{xz}^{(1)} | \widehat{H} | d_{xz}^{(1)} \rangle$ & $(0,0,0)$ & $10.66$ & $10.69$ \\
$\varepsilon_{d_{xy}}^{(1)}$ & $\langle d_{xy}^{(1)} | \widehat{H} | d_{xy}^{(1)} \rangle$ & $(0,0,0)$ & $10.69$ & $10.68$ \\
$\varepsilon_{d_{xz}}^{(2)}$ & $\langle d_{xz}^{(2)} | \widehat{H} | d_{xz}^{(2)} \rangle$ & $(0,0,0)$ & $10.66$ & $10.69$ \\
$\varepsilon_{d_{xy}}^{(2)}$ & $\langle d_{xy}^{(2)} | \widehat{H} | d_{xy}^{(2)} \rangle$ & $(0,0,0)$ & $10.69$ & $10.68$ \\
$\Delta\varepsilon^{(1)}$ & $\varepsilon_{d_{xz}}^{(1)} - \varepsilon_{d_{xy}}^{(1)}$ & $(0,0,0)$ & $0.03$ & $0.01$ \\
$\Delta\varepsilon^{(2)}$ & $\varepsilon_{d_{xz}}^{(2)} - \varepsilon_{d_{xy}}^{(2)}$ & $(0,0,0)$ & $0.03$ & $0.01$ \\
\bottomrule
\end{tabular}
\begin{tablenotes}\footnotesize
\item[*] The numbers $(1)$ and $(2)$ indicate the two inequivalent Hf sites (Hf1 and Hf2) in the buckled honeycomb layer.
\end{tablenotes}
\end{threeparttable}
\end{table}

Fig.~\ref{figure6} compares the imaginary part of the second-nearest-neighbor hopping matrix $\mathrm{Im}(t_2)$ for the low-symmetry $P1$ structure and the higher-symmetry $P321$ case grouped by orbital indices $(i, j)$. These imaginary components are crucial indicators of spin-orbit-induced complex phases and inter-orbital hybridization. In the $P1$ phase, several matrix elements such as $t_{2,3}$, $t_{1,4}$ and their transposes show large non-zero imaginary contributions breaking Hermitian symmetry. This reflects enhanced inter-orbital hybridization and spin-orbit entanglement, enabled by the absence of inversion symmetry. The asymmetry in color distribution across $(i,j)$ and $(j,i)$ further highlights broken inversion and the emergence of Rashba-type effects with $t_2 $ amplitudes reaching up to 20 meV. Due to preserved inversion symmetry, the $P321$ matrix exhibits minimal imaginary components across all orbital pairs below 4 meV and is nearly Hermitian. The heatmap structure appears more symmetric suggesting that additional crystalline symmetry constraints suppress complex SOC-mediated hopping paths. The near-zero off-diagonal elements imply minimal inter-orbital coupling, which aligns with weaker SOC-induced Berry curvature and diminished Rashba splitting. Altogether, these heatmaps exemplify how the imaginary part of $t_2$ provides a direct microscopic fingerprint of Rashba-active symmetry breaking. The comparison also shows that $\sum |\mathrm{Im}(t_{ij}) + \mathrm{Im}(t_{ji})|$ serves as a sensitive indicator for identifying SOC-induced inversion asymmetry and is maximized in the non-centrosymmetric $P1$ phase, while high-symmetry phases like $P321$ enforce constraints that dampen these features. 

The emergence of non-zero imaginary inter-orbital hopping elements in the $P1$ phase echoes the general mechanism described by Zunger \textit{et al.}~\cite{mera2020rashba}, where Rashba splitting arises from SOC-induced hybridization between orbitals of different symmetry in a non-centrosymmetric environment. In their one-dimensional model, the combination of spin-orbit coupling and orbital asymmetry gives rise to complex-valued hopping terms, which in turn drive spin-momentum locking. Our two-dimensional tight-binding Hamiltonian based on $d_{xy}$ and $d_{xz}$ orbitals reproduces this scenario in the context of oxide heterostructures. In particular, the enhanced imaginary components of $t_2$ in the $P1$ structure serve as microscopic indicators of SOC-driven orbital mixing as described by the Rashba mechanism in their analytical model. This is further supported by the near-degeneracy of the on-site orbital energies shown in Table~\ref{tab:P1_P321_TB_Comparison}, where the $d_{xy}$ and $d_{xz}$ states on Hf1 and Hf2 differ by less than 35~meV in $P1$, compared to more pronounced splittings and suppressed imaginary hoppings in the centrosymmetric $P321$ phase. While such near-degeneracy could, in principle, support strong Rashba splitting via band anti-crossing, Fig.~\ref{figure2}b confirms that no crossing exists in the spin-degenerate limit. As a result, the Rashba effect in our system arises from SOC acting on already non-degenerate bands, placing it in the type II Rashba category in the classification of Zunger \textit{et al.}~\cite{mera2020rashba}. This highlights that spin-orbit-induced orbital hybridization and symmetry breaking can generate moderate Rashba splitting even in the absence of band crossings. This orbital-level interpretation of Rashba splitting is further substantiated by the work of Szary~\cite{szary2023rashba}, who highlights the central role of unquenched orbital angular momentum (OAM) and symmetry-allowed orbital hybridization. In his chemical physicist’s approach, Rashba effects arise from SOC-induced mixing between orthogonal orbitals (e.g., $p_x$/$p_y$, $d_{x^2 - y^2}$/$d_{xy}$) under inversion symmetry breaking. Our system, featuring hybridized $d_{xy}$ and $d_{xz}$ orbitals in the low-symmetry $P1$ phase, fulfills these criteria. Such orbital combinations sharing the same total angular momentum quantum number $l$ can still generate $k$-antisymmetric OAM when inversion symmetry is broken. The observed complex hopping amplitudes such as $\mathrm{Im}(t_2)$ therefore serve as a fingerprint of this mechanism, which is in accordance with Szary’s orbital perspective and supporting the microscopic origin of the Rashba effect in our oxide heterostructure.

\begin{figure} [htbp!]
\centering
\includegraphics[width=0.5\textwidth]{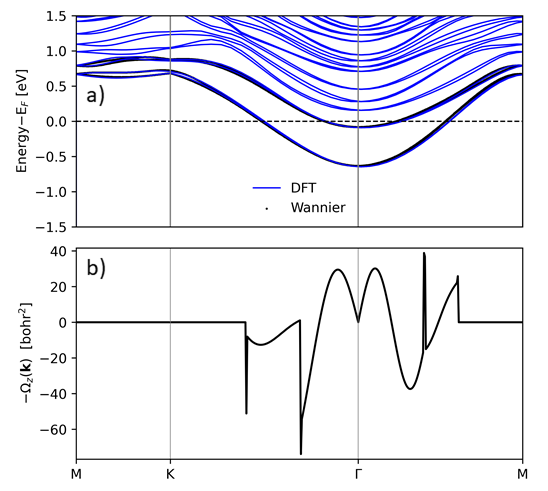}
\caption{a) Comparison between Wannier-interpolated (black) and DFT (blue) band structures (top) and b) the corresponding Berry curvature (bottom) along the high-symmetry path $M \rightarrow K \rightarrow \Gamma \rightarrow M$. While Rashba-type spin splitting is clearly visible along the $M \rightarrow K$ path, the Berry curvature is strongly peaked near the $\Gamma$-point. This contrast highlights the geometric origin of Berry curvature, which arises from SOC-induced avoided crossings and wavefunction mixing near band extrema, rather than from spin splitting alone.}
\label{figure7}
\end{figure}

Fig.~\ref{figure7}a displays the excellent agreement between DFT and Wannier-projected bands in the energy window near the Fermi level, confirming the fidelity of the Wannier representation and thereby validating the use of the interpolated model in analyzing Berry curvature. These avoided crossings are accompanied by significant wavefunction mixing, which in turn leads to large Berry curvature responses as seen in Fig.~\ref{figure7}b. In contrast, the $M$--$K$ region exhibits pronounced Rashba spin splitting due to inversion asymmetry and SOC, yet contributes little to the Berry curvature. This is attributed to the absence of nearby band mixing that suppresses curvature generation along this high-symmetry line. This analysis provides a direct link between SOC, avoided crossings and Berry curvature generation. It supports the notion that Berry curvature is a geometric property tied to wavefunction entanglement, rather than a purely energetic quantity like spin splitting, i.e., the emergence of Berry curvature even in systems with vanishing anomalous Hall conductivity. 

Inter-orbital mixing refers to non-zero hopping matrix elements of the form $t' = \langle \phi_i | \widehat{H} | \phi_j \rangle$ for $i \neq j$,
where $\phi_i$ and $\phi_j$ are different orbitals (e.g., $d_{xy}$, $d_{xz}$) or spin-mixed states (e.g., $d_{xz}^\uparrow$, $d_{xz}^\downarrow$). In the $P1$ phase, the inter-orbital hopping terms such as $t'$ and $t''$ reach values up to $\sim 29-30$\,meV (real part), while in the $P321$ phase they are significantly suppressed to $\sim 7-9$\,meV due to the higher crystalline symmetry. This enhanced $t'$ in the $P1$ phase facilitates stronger orbital hybridization under SOC, promoting larger avoided crossings and stronger momentum-dependent band repulsion. As a result, the spin splitting becomes more pronounced and the Berry curvature near the band extrema (see Fig.~\ref{figure7}b) is amplified with $-\Omega_z(\mathbf{k})$ reaching values below $-60$ and above $+40$ bohr$^{-2}$ around the $\Gamma$-point. However, because these peaks exhibit alternating signs, the integrated Berry curvature along the high-symmetry path remains nearly zero due to cancellation. In stark contrast, the $P321$ structure with its trigonal symmetry and threefold rotation axis restricts many off-diagonal hopping terms by suppressing inter-orbital mixing. Consequently, the SOC-induced mixing is weakened, leading to less pronounced Rashba splitting and reduced Berry curvature. The takeaway is that the Rashba effect in these oxide heterostructures is not only enabled by SOC and inversion symmetry breaking, but also magnified by the degree of inter-orbital hybridization. Thus, the stronger $t'$ in $P1$ leads to more pronounced spin splitting, larger Berry curvature hotspots and potentially improved spin-charge conversion efficiency in applications such as spintronics.

\section{Summary}

In this work, we systematically investigated the Rashba spin splitting in buckled honeycomb oxide heterostructures based on the (SrHfO$_3$)$_2$/(LaAlO$_3$)$_4$(111) superlattice using a combination of DFT, tight-binding modeling and symmetry analysis. Our first-principles calculations within GGA\,+\,$U$\,+\,SOC confirm the presence of a spin-split metallic state in the non-centrosymmetric $P1$ phase, characterized by a clear Rashba-type band splitting near the $M$ and $K$ points and a chiral in-plane spin texture and finite out-of-plane polarization. In contrast, the higher-symmetry $P321$ phase remains spin-degenerate, consistent with inversion and rotational symmetry constraints. To analyze the origin of these features, a minimal tight-binding Hamiltonian has been constructed using maximally localized Wannier functions projected onto the Hf 5$d_{xy}$ and 5$d_{xz}$ orbitals. SOC was incorporated analytically via an atomic $\lambda \vec{L} \cdot \vec{S}$ term allowing interpolation of $\lambda$ to probe the onset of spin splitting. The model captures the essential Rashba characteristics and reproduces the DFT band dispersion and spin-momentum locking. We quantified the Rashba spin splitting by two complementary approaches: (i) a direct DFT-based extraction of the Rashba energy $E_R$ and momentum offset $k_R$ from the DFT band structure near the $K$ point yielding $E_R = 29$\,meV and $k_R = 0.17$\,\AA$^{-1}$, which correspond to a Rashba coefficient of $\alpha_R = 0.34$\,eV$\cdot$\AA\ and (ii) a $k \cdot p$ theory-based fit to the two lowest spin-split TB bands at fixed $\lambda_{\rm SOC} = 0.15$\,eV resulting in $\alpha_R = 0.33$\,eV$\cdot$\AA. The close agreement between both methods corroborates the reliability of the effective Rashba model in capturing low-energy spin-splitting phenomena. Furthermore, $\Gamma$-point phonon calculation including SOC confirmed the dynamical stability of the $P1$ phase. Analysis of the TB matrix elements revealed that inversion symmetry breaking in $P1$ introduces enhanced imaginary inter-orbital hopping terms, which correlate with increased spin splitting and Berry curvature near the Fermi level. In summary, our results highlight the crucial role of symmetry, orbital mixing and atomic SOC in enabling Rashba-type spin-orbit effects in oxide heterostructures. The SrHfO$_3$-based buckled honeycomb system provides a tunable platform for exploring spin textures, symmetry-protected band degeneracies and spin-orbit-enabled functionalities in oxide spintronics.


\section*{DATA AVAILABILITY}

The author declares that the main data supporting the findings of this study are available within the article and its Supplementary Information files. Additional data can be provided upon reasonable request.

\section*{ACKNOWLEDGMENTS}

I gratefully acknowledge the computational resources (Stevin Supercomputer Infrastructure) and services used in this work were provided by the VSC (Flemish Supercomputer Center), funded by Ghent University, FWO and the Flemish Government - department EWI.


\section*{COMPETING INTERESTS}

The author declares no competing financial or non-financial interests. 


\bibliographystyle{apsrev4-2}
\bibliography{references}

\end{document}